\documentclass[conference]{IEEEtran}
\usepackage{cite}
\usepackage{amsmath,amssymb,amsfonts}
\usepackage{algorithmic}
\usepackage[bookmarks=false]{hyperref}
\usepackage{listings}
\usepackage{graphicx}
\usepackage{xcolor}
\usepackage{tabularx}
\usepackage{tabulary}
\usepackage{array}
\usepackage{booktabs}
\usepackage{pifont}             
\usepackage[shortlabels]{enumitem}			
\usepackage{makecell} 

\usepackage{xcolor} 
\definecolor{light-gray}{gray}{0.95}

\begin{document}

\title{Modeling and Executing Production Processes with Capabilities and Skills using Ontologies and BPMN}

\author{
\IEEEauthorblockN{
    Aljosha Köcher\IEEEauthorrefmark{1},
    Luis Miguel Vieira da Silva\IEEEauthorrefmark{1},
    Alexander Fay\IEEEauthorrefmark{1},
}
\IEEEauthorblockA{
\IEEEauthorrefmark{1}Institute of Automation\\
Helmut Schmidt University, Hamburg, Germany\\
Email: aljosha.koecher@hsu-hh.de, miguel.vieira@hsu-hh.de, alexander.fay@hsu-hh.de\\}
}

\maketitle

\begin{abstract}
Current challenges of the manufacturing industry require modular and changeable manufacturing systems that can be adapted to variable conditions with little effort. At the same time, production recipes typically represent important company know-how that should not be directly tied to changing plant configurations.
Thus, there is a need to model general production recipes independent of specific plant layouts. For execution of such a recipe however, a binding to then available production resources needs to be made.
In this contribution, we select a suitable modeling language to model and execute such recipes.
Furthermore, we present an approach to solve the issue of recipe modeling and execution in modular plants using semantically modeled capabilities and skills as well as BPMN. 
We make use of BPMN to model production recipes using \emph{capability processes}, i.e. production processes referencing abstract descriptions of resource functions. 
These capability processes are not bound to a certain plant layout, as there can be multiple resources fulfilling the same capability. 
For execution, every capability in a capability process is replaced by a skill realizing it, effectively creating a \emph{skill process} consisting of various skill invocations.
The presented solution is capable of orchestrating and executing complex processes that integrate production steps with typical IT functionalities such as error handling, user interactions and notifications. Benefits of the approach are demonstrated using a flexible manufacturing system.
\end{abstract}


\section{Introduction}
\label{sec:Introduction}
The current situation in which manufacturing companies operate is characterized by increasing fluctuations of both customer demands and supply of raw materials. In such an environment, it is necessary to have a high potential to adapt to changes in order to be successful. 
In manufacturing, the degree of this so-called adaptability is mostly defined by the way that production equipment can be reconfigured for different variations of products and to react on fluctuations of customer demands \cite{VDI_5201_Descriptionandmeasurementof}.

The term \emph{Plug \& Produce} has been coined to describe a simple and quick method of integrating new manufacturing machines and their functions into a plant without tedious manual configuration or reprogramming efforts \cite{AAM+_AgileAssemblySystemby_2000}. A description of manufacturing modules and their functionalities is vital for the concept of Plug \& Produce in order for module functions to be used by higher-level systems.

In \cite{KHV+_AFormalCapabilityand_9820209112020}, we have provided a model of machines, their capabilities and skills in the form of an ontology which consists of separate so-called \emph{ontology design patterns} (ODPs) that are all based on industry standards.
According to \cite{Jar_CapabilitybasedAdaptationofProduction_2012}, capabilities can be seen as the processes a machine is able to perform. Skills are considered to be the executable counterpart of capabilities, i.e. a skill is a technical implementation realizing a capability. Every skill provides an interface description so that it can be invoked and monitored from other systems, e.g., manufacturing execution systems (MES) \cite{KHV+_AFormalCapabilityand_9820209112020}.
In addition to the model presented in \cite{KHV+_AFormalCapabilityand_9820209112020}, we have previously presented automated methods to create module and capability descriptions as well as skills in \cite{KHC+_AutomatingtheDevelopmentof_9820209112020} and \cite{ KJF_AMethodtoAutomatically_2021}. These methods allow to register skills implemented in either IEC 61131 or Java and that are executable via OPC UA or as web services in a skill-based MES. Using these methods, a plant can be put together in a dynamic way using semantically described modules, their capabilities and their skills.

In contrast to this dynamicity of modules and skills, production recipes describing the process steps for manufacturing a product are much less dynamic as they encode a company's production knowledge.
IEC 61512 defines four different types of recipes, of which we will only use the most abstract and most concrete one: While a \emph{general recipe} is independent of any location and plant, a \emph{control recipe} is tailored to a specific lot and contains all relevant control information (e.g., equipment used and specific parameters) \cite{IEC_615121_Batchcontrol}. 
With this contribution, we aim to present an approach to digitally model general production recipes in a reusable way and be able to convert them to control recipes that allow for recipe execution under varying plant layouts. To achieve this, the following research questions are answered:
\begin{itemize}
    \item Which existing modeling languages may be used for production recipes and how can these existing languages be evaluated in order to find the best suited process modeling language for capability- and skill-based production?
    \item How can capabilities and skills, which are defined in an ontology, be used when modeling production recipes? 
    \item How can production recipes be integrated with necessary utility functions such as error handling, user interaction and notification?
\end{itemize}

It's important to note that this paper is no contribution to the topic of \emph{capability matching}, i.e., finding provided capabilities that are able to fulfill a set of required ones as discussed in \cite{Jarvenpaa.2017}. Instead, we aim to provide an orchestration language and execution mechanism for sequences of provided capabilities. Whether these sequences are manually modeled - as is the case for conventional production planning - or found by means of a previous capability matching procedure, does not matter.

The remainder of this paper begins with the selection of a process modeling language that can be used for production recipes in the context of capabilities and skills. Our own approach is presented in Section \ref{sec:approach} and evaluated in Section \ref{sec:evaluation}. Afterwards in Section \ref{sec:relatedWork}, previous approaches to model and execute production processes are discussed and compared with the approach presented in this contribution before a summary and outlook on future research in Section \ref{sec:summary}.

\section{Selection of a Process Modeling Language}
\label{sec:processModellingLanguages}
In order to model production recipes using capabilities and later execute them as skills, a suitable modeling language is required.
This modeling language should fulfill four substantial requirements, some of which were taken from VDI guideline 3681, a guideline for comparing modeling languages \cite{VDIVDE_3681_Classificationandevaluationof}:
\begin{itemize}
    \item The modeling language must have a formal grounding, which permits to model complex processes without any ambiguity in interpretation.
    \item A serialization format must exist - ideally in a standardized way, so that modeled processes can be used by software tools and exchanged between tools if necessary.
    \item The modeling language must be able to represent separate capabilities that can be executed as skills in the sense of an orchestration language.
    \item There must be existing software tools - ideally for both modeling and execution of processes, because developing a new modeling and execution software from scratch is out of scope of this work.
\end{itemize}

When searching for a suitable modeling language, we looked at \emph{Formalized Process Description} (FPD) as defined in VDI guideline 3682, finite state machines (FSM), Petri nets and \emph{Business Process Model and Notation} (BPMN).

Since we model individual capabilities as process operators of VDI 3682, there was the consideration to model sequences of several capabilities as processes of VDI 3682. However, this guideline does not meet any of the above criteria. The process model is not sufficiently and not formally enough defined, so that there are ambiguities, especially in the case of alternative or parallel flows. In addition, there is currently no standardized modeling format and with \cite{NKH+_OffeneswebbasiertesWerkzeugzur_2020}, there is currently only one software tool, which is only capable of  modeling processes. An execution of processes modeled according to VDI 3682 is currently not possible.

More formally defined and widespread than VDI 3682 are FSM and Petri nets. Both meet the criteria for a high degree of formalization and both provide textual representations that can be used for data exchange. Furthermore, there are various software tools for both modeling and analysis. However, there are no tools that support FSM or Petri nets as general orchestration languages: We are not aware of any tool that would allow modeling processes which consist of individual calls to different skills using FSM or Petri nets.

BPMN fulfills all the requirements above. It provides formalized execution semantics, and to some extent BPMN resembles Petri nets and is thus used for formal verification \cite{CFP+_Aformalapproachfor_2021}. BPMN has a serialization format standardized by the Object Management Group \cite{OMG_BPMN_BusinessProcessModeland} and there is a large variety of tools for modeling and executing BPMN processes. With BPMN, processes consist of different types of tasks which can be used to represent all kinds of external functionality - such as skill invocations in our case. Because all requirements are met and especially because there are tools for modeling processes as well as so-called \emph{execution engines} that can automatically execute BPMN processes, BPMN is chosen.
An approach to model processes based on capabilities and execute these processes using skills is presented in the following section.

\section{Modeling and Executing Skills using BPMN}
\label{sec:approach}
In the following two subsections, an approach to model and execute capability processes using BPMN is presented. 
The approach is based on our ontological capability and skill model. Thus, the essential elements of this capability and skill model are introduced in Subsection \ref{subsec:capabilityModel}. Subsection \ref{subsec:modelling} then describes how general production recipes can be modeled using capabilities and how necessary information is integrated into a standard-conformant BPMN model. Subsection \ref{subsec:execution} shows how such BPMN models are transformed into executable processes by swapping capabilities against skills that are available in a given manufacturing layout.

\subsection{Capability and Skill Model}
\label{subsec:capabilityModel}

\begin{figure*}[h]
    \centering
    \includegraphics[width=0.9\linewidth]{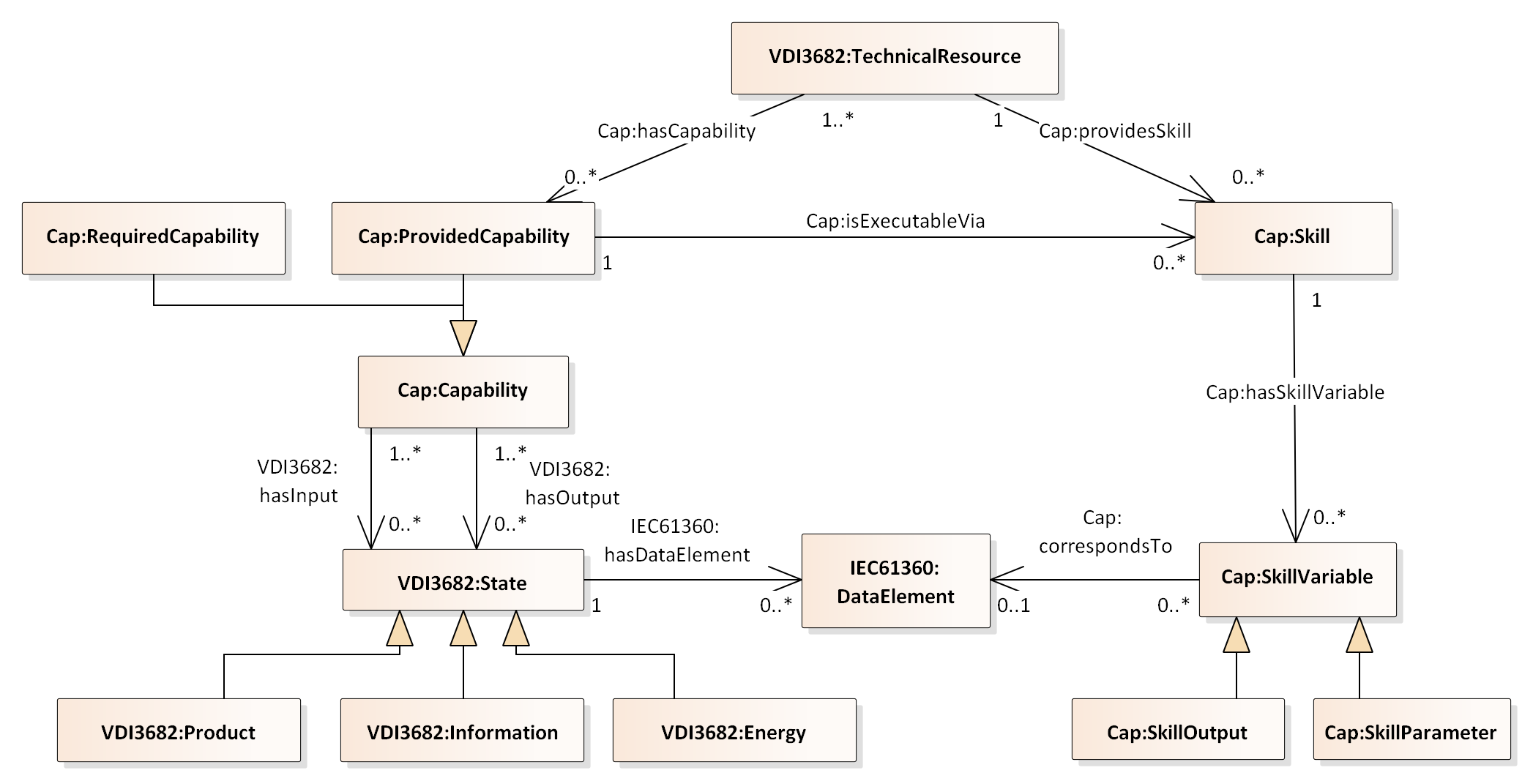}
    \caption{An excerpt of the capability and skill ontology presented in \cite{KHV+_AFormalCapabilityand_9820209112020} that is most relevant for this contribution. While one provided capability may be offered by multiple machines, every machine typically has one or more unique skills to execute this capability.}
    \label{fig:ontologySnippet}
\end{figure*}

In our previous publication \cite{KHV+_AFormalCapabilityand_9820209112020}, a capability and skill model was introduced in the form of an ontology consisting of different standard-compliant ontology design patterns. Of particular relevance to this paper is the way that capabilities and skills are related to each other (see Figure~\ref{fig:ontologySnippet}). In Figure~\ref{fig:ontologySnippet}, prefixes such as \emph{VDI3682} correspond to a standard-compliant ontology while the prefix \emph{Cap} highlights model elements connecting the standard-compliant model elements to one joint capability and skill model.

In this model, required capabilities are distinguished from provided ones. Required capabilities are used by external entities to request certain processes or products (e.g. via a marketplace). Required capabilities need to be matched against provided ones. 
This aspect is out of scope of this contribution because it focuses on internal production planning with given machines and their provided capabilities.
One provided capability may be offered by multiple machines. Skills on the other hand are unique to a machine. Every machine might provide one or more skills to execute certain capabilities.
Capabilities are modeled as process operators according to \cite{VDI_36821_Formalisedprocessdescriptions} that can have input and output state elements (products, energy, information). These elements can in turn have properties which are described with data elements according to \cite{IEC_61360_Standarddataelementtypes}. These data elements are properties that allow to describe constraints about the possible inputs and outputs of capabilities. Such properties can be used, for example, to express that a drilling capability can only drill holes to a certain depth.

A skill is the technical implementation of a capability with an invocation interface (e.g. using OPC UA). Our ontological model contains a description of skills and their invocation interfaces so that other systems can use this description to interact with a skill. Skills follow a state machine according to \cite{ANSIISA_TR88.00.022015_MachineandUnitStates}. 
An invocation of a skill is no synchronous function call, but instead it is an asynchronous transition of the skill into a different state depending on the current state. Actions are performed during states. The main skill action is typically performed during the \emph{execute} state, while e.g. error-handling is done during \emph{stopping} or \emph{aborting}.
In order to trigger transitions, skills provide corresponding methods. And in order to configure a skills behavior, it has skill variables --- which can either be parameters or result outputs. 
Skill variables may be linked to the associated capability data elements. For example, the data element used to express a constraint on the depth of a hole can be linked to a skill parameter, which allows setting the desired drilling depth.

\subsection{Capability Process Modeling Approach}
\label{subsec:modelling}
In order to describe general production recipes according to IEC 61512 as processes consisting of multiple capabilities, we make use of BPMN.
BPMN does not provide any builtin model elements to represent capabilities or skills, so an approach to represent these with the standardized model elements is required. In BPMN, processes consist of various activities which can be either atomic (so-called \emph{tasks}) or compound (so-called \emph{sub-processes}). 
Tasks are subdivided into seven different task types: \emph{Send Tasks} and \emph{Receive Tasks} allow to send and receive messages to / from external entities, respectively. Human interaction can be modeled using a \emph{User Task} or a \emph{Manual Task}. While the first one is tracked by a BPMN engine, tasks of the latter type are to be performed without an engine. Automation of processes can be achieved through \emph{Script Tasks}, \emph{Business Rule Tasks} or \emph{Service Tasks}. 
Script Tasks are used to implement smaller functionalities which are executed directly by a BPMN engine. 
Business Rule Tasks define rules that are checked by a separate rule engine. 
Service Tasks offer the greatest potential for modeling capabilities because they allow delegation of a task to an external application which will be in charge of executing it.
As soon as a process execution reaches a service task, the external application referenced by the Service Task is called. 
When that external application returns, the BPMN engine continues the execution of the remaining activities of the process. 
This pattern of interaction allows to extend basic BPMN execution by creating external applications handling specific aspects. 

Due to this possibility of extending BPMN, service tasks are used to model so-called \emph{capability processes}, i.e. BPMN processes with capabilities referenced through service tasks. As capabilities are only a description of machine functions, these capability processes need to be converted to \emph{skill processes} at execution time so that they can be automatically executed by a BPMN engine.

\smallskip

For our approach to model capability-processes, the open-source BPMN modeling tool \emph{bpmn-js}\footnote{https://github.com/bpmn-io/bpmn-js} was extended and integrated into the skill-based manufacturing execution system \emph{SkillMEx}\footnote{https://github.com/aljoshakoecher/SkillMEx} to have access to all registered machines with their capabilities and skills. 
A user can model BPMN processes using all conventional types of BPMN elements. Service Tasks were extended so that they can represent usage of a provided capability in a process. For each service task referencing a capability, a skill linked to this capability is later selected in order to execute the process.

After adding a service task element to a process, a user can select a capability from the ones that are currently registered in SkillMEx. Required input properties of the selected capability are loaded and can be set with either constant values or variables.
If a constant value is used for a capability property, this value is used for every execution of this process independent of the concrete skill that will later replace this capability.
Variable capability property values can be used to defer binding of concrete property values to the process execution stage. During execution, variable values may be calculated or set with a user task that requests user input for a variable value.
Output properties are added into the process model and can be used as inputs of subsequent tasks or in conditions (e.g. of alternative flows). 
When a capability is chosen in a service task, this capability along with its input and output properties and their values is stored inside the definition of this task and may be serialized as BPMN-compliant XML.

In addition to using tasks, advanced BPMN elements such as timers or error boundary events may be used as well. Error events allow to handle errors that may later arise during skill execution e.g. with dedicated tasks or a sub process.
An example of a process definition along with the user interface to add a capability inside a service task is presented in Figure~\ref{fig:evaluation-example}. 

After modeling of a capability process is complete, the process model can be deployed to a BPMN engine. For execution, every capability in a capability process needs to be resolved to a skill effectively creating a skill process. Such skill processes can be compared to control recipes according to IEC 61512.
How this resolving works and how such skill processes are ultimately executed through a BPMN engine is the subject of the following subsection.

\subsection{Process Execution}
\label{subsec:execution}
In order to execute BPMN processes containing capabilities using SkillMEx, we connected Camunda's open-source BPMN engine\footnote{https://github.com/camunda/camunda-bpm-platform} to it. After a process has been modeled, it is stored with its complete XML representation which contains all capabilities and properties defined during modeling.
As soon as a user wants to start a process, the selected process definition is loaded from the BPMN engine's database. For every capability defined in the process definition, all currently available skills are retrieved from SkillMEx' ontology. In case there is more than one skill for a capability, a user selects the skill to be used. 
As skill parameters are directly linked to capability properties (compare Fig~\ref{fig:ontologySnippet}), property values of the process definition are mapped to skill parameters. After this mapping of capabilities to skills, the process is ready to be executed and can be started.

\begin{figure*}
    \centering
    \includegraphics[width=0.9\textwidth]{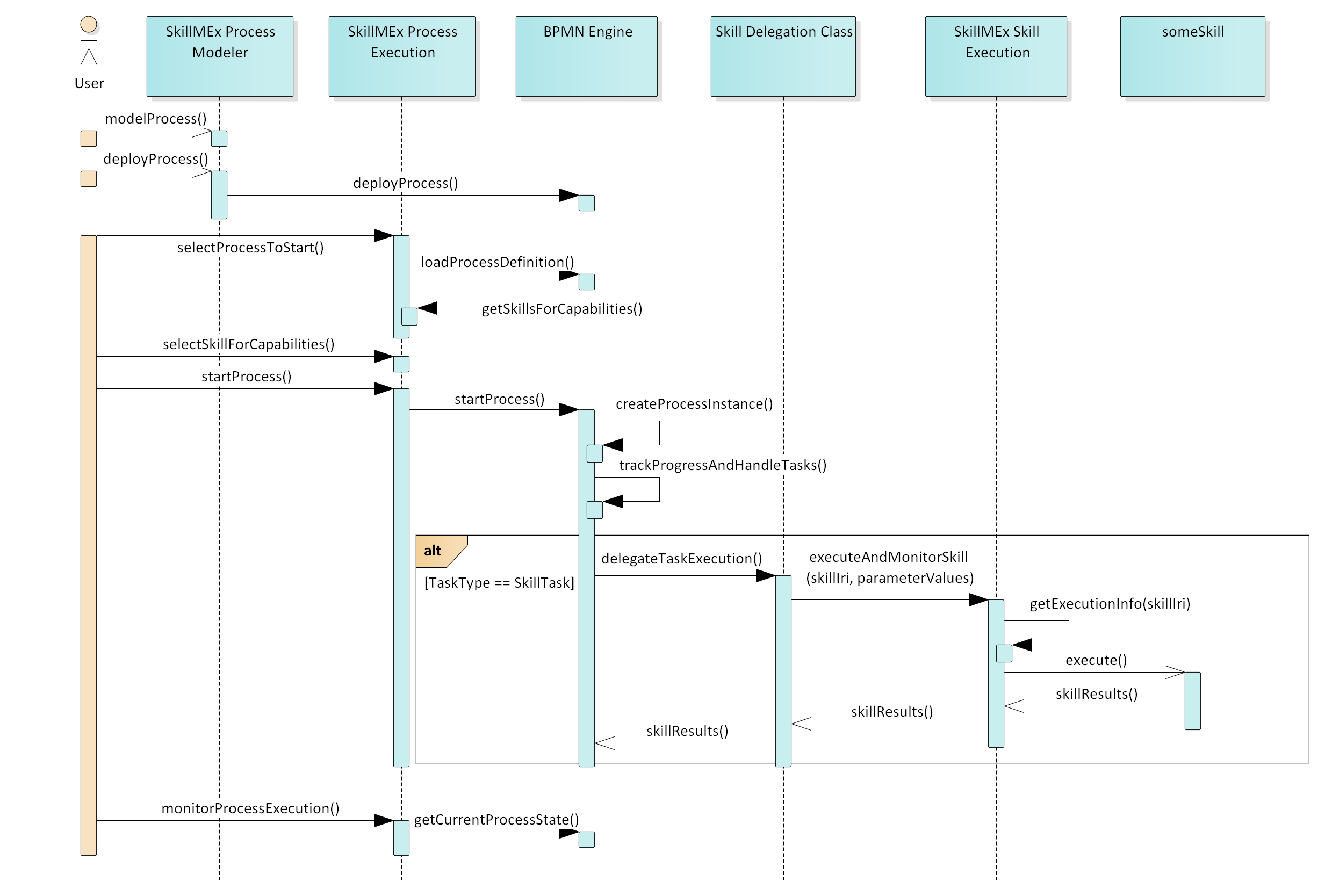}
    \caption{Systems and the sequence required to execute BPMN processes containing skills. Note that \emph{someModule} is a placeholder for any resource with skills.}
    \label{fig:skillExecution}
\end{figure*}

\smallskip
In order to execute a process, a new process instance is created by the BPMN engine. Beginning with the start event, the engine keeps track of the process state and executes all tasks in the modeled order. 
As soon as a Service Task with a skill definition is encountered, execution of this task is delegated to a custom delegation class\footnote{https://github.com/aljoshakoecher/BPMN-Skill-Executor} which we developed as a way to extend the default behavior of the Camunda BPMN engine. All information required for invocation of a skill such as its unique \emph{IRI}, the transition to trigger and all parameter values are passed to this class.
This delegation class in turn uses a skill execution interface from SkillMEx to call skills. 
This skill execution interface is able to determine all technical detail information required for the actual invocation of a skill provided by a specific resource. This might include the protocol used, the (IP) address of a skill and the way in which a skill call is made and parameters are transferred. 
For example, in this step, the OPC UA server of a skill is determined and the specified transition of a skill is executed via an OPC UA method call. Interaction with a skill is asynchronous, i.e. the effect of a transition can last for an indeterminate amount of time and the skill execution interface does not wait for feedback. Instead, after invoking a transition on a skill, its state changes are monitored by the delegation class. For this purpose, every skill publishes its current state or provides means for the delegation class to get the current state.

Once the \emph{Complete} state is reached, all skill outputs are returned to the BPMN process execution so that they are available for subsequent steps. Furthermore, the task is marked as completed. The engine then continues process execution.
Instead of successfully completing, a skill might also end up in an error state such as \emph{Stopped} or \emph{Aborted}. In such cases, the delegation class throws errors back to the process engine so that error handlers of the BPMN process are triggered. This allows for transparent error handling to be modeled in a BPMN process definition.

At any time of process execution, a user may monitor progress of a process and all current parameter values. Camunda's BPMN engine provides an interface to load current process information which is displayed in SkillMEx.
The complete flow used to execute a BPMN process containing skills of machines is depicted in Figure~\ref{fig:skillExecution}. Please note that in this figure \emph{someSkill} is just a placeholder for a skill. In reality, there can of course be multiple skills of various machines involved in a process execution.

\section{Evaluation}
\label{sec:evaluation}
In order to evaluate our approach, we applied it to a \emph{Festo MPS 500}, a lab-scale modular production system. We make use of three different module types: A \emph{Thermometer Supply Module} able to store and supply the base material for small thermometers and a \emph{Drilling Module} are connected via a \emph{TransportModule}. While there is only one instance of the Thermometer Supply Module and the Transport Module, there are two physical instances of the Drilling Module. 

\begin{figure*}[h]
    \centering
    \includegraphics[width=0.8\textwidth]{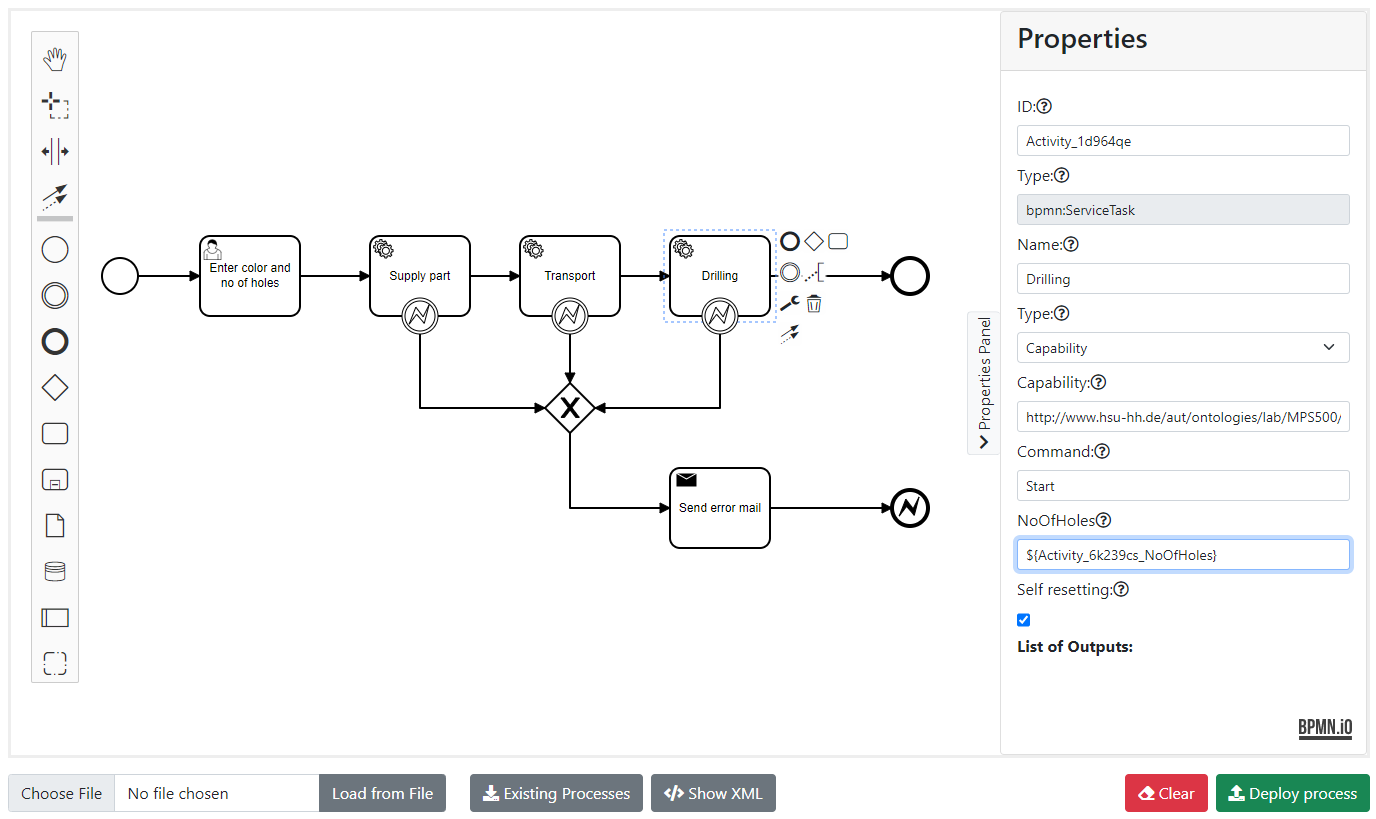}
    \caption{Exemplary BPMN process model used to test the approach on a modular plant. A user input ("\$\{Activity\_6k239cs\_NoOfHoles\}") is used to parameterize the drilling capability (see properties panel on the right).}
    \label{fig:evaluation-example}
\end{figure*}

We start with a scenario in which the Thermometer Supply Module is connected to one Drilling Module via a conveyor belt of the Transport Module. The production process we consider is shown in Figure~\ref{fig:evaluation-example}: First, a user task requires a user to select a thermometer color and to enter the number of holes that are to be drilled. These are arbitrary inputs that can be used to create different product configurations.
In this process, a user task is used to set values of capability properties at process execution time. If constant values were set, they would be fixed for every execution of the process. This can be useful for overall configuration properties which need to have the same value for every process execution instance but not for product customization.
After the user task, the three capabilities \emph{Supply Part}, \emph{Transport} and \emph{Drilling} are modeled in a sequence. See Figure~\ref{fig:evaluation-example} for a screenshot of the process modeled inside our skill-based MES. The properties panel on the right side of the figure shows how capabilities are set to a BPMN service task turning them to capability tasks. The parameter \emph{\$\{Activity\_6k239cs\_NoOfHoles\}}, which is an output of the user task, is used as a value for the \emph{Drilling} that is currently selected.

In addition to the planned sequence of capabilities, all capability tasks are equipped with a so-called error boundary event, which catches errors during later skill execution. These events direct the control flow over a mail send task to an alternative end of the process. The mail send task is used to inform responsible recipients e.g., plant operation personnel. In the process shown in Figure~\ref{fig:evaluation-example}, an email with a short notification is automatically sent.

To execute this process, all skills that are available for every capability of the process are determined. For this example process, there is only one skill for each capability so that the execution of the process can be started directly.

The Drilling Skill is implemented in a way that an error is raised as soon as a part is removed during the process (e.g. by manual intervention). We purposefully trigger this error to simulate a failure of the module. Switching to an error state during skill execution triggers the error catch event of the drilling skill so that the \emph{error control flow} is started and a notification is sent.

This (simulated) error is taken as a reason to replace the currently available drilling module with the second one.
Important to note is that the previously modeled production process can be reused without any changes because it was modeled based on capabilities and both drilling modules offer the same capability. The only change is that for execution, the skill of the new drilling module is now chosen to realize the drilling capability with its skill.

If both modules are available in the plant at the same time, the process would also be executable. In this case, one of the two skills would have to be selected manually to realize the drilling capability.

We recognize that our current evaluation is an early test to prove the overall applicability of the approach and to foster further development.
Due to this early evaluation stage, it is difficult to provide quantifiable data, e.g. on time saved. One advantage of this approach that can already be noted is that process models can be reused without having to change anything if skills are changed.
Compared to approaches that directly link process models to executable skills, the presented approach eliminates the need to re-model or change process models just because other or additional skills are available for certain capabilities.
This aspect is not only important because it reduces modeling efforts, but also because it allows to reuse established and tested processes. In addition to modeling efforts, changing established processes could lead to time-intensive test runs.

\section{Related Work}
\label{sec:relatedWork}
Our approach meets the criteria of \emph{orchestration} in the context of manufacturing automation as given by Stutz et al. \cite{SFB+_Orchestrationvs.ChoreographyFunctional_2020}. Therefore, we present an overview of existing approaches to orchestration in this section. As this is a rather broad field of research, we focus on approaches of manufacturing automation that use capabilities, skills or similar kinds of encapsulating and describing machine functions. For a more general overview on web service orchestration, we refer to \cite{HuSi_Compositionofheterogeneousweb_2019}


A rather early contribution to orchestration of automated machine functions is presented in \cite{THJ_ServiceorchestrationwithOPC_2014}. The authors developed a graphical modeling language similar to Sequential Function Charts that is used to orchestrate OPC UA devices. OPC UA methods can be invoked to execute a given behavior and OPC UA variables can be used as guard conditions. 
The fact that the modeling language is a custom-developed language and not a standardized one and that it is tailored to OPC UA are major drawbacks. The approach by \cite{THJ_ServiceorchestrationwithOPC_2014} is not applicable for an application in which skills can be invoked using different technologies and in which further functionalities (such as human intervention or notifications) need to be integrated.

Dorofeev et al. present an approach to generate orchestration code in the context of skill-based automation. In order to model and implement individual skills, function blocks according to IEC 61499 are used. Based on these single skills, a user can provide a textual encoding of an orchestration, i.e. a sequence of skill invocations. The textual specification of skill sequences is a self-developed language that does not provide the features of standardized and widely used languages such as BPMN. Furthermore, sequences are directly bound to individual skills, so there is no abstraction (e.g. on the basis of capabilities) that would allow reuse of process models after reconfiguring a plant \cite{DBT+_GenerationoftheOrchestrator_9720219102021}.

In \cite{RiVo_ModelingofManufacturingExecution_2010}, the authors show a comparison of different modeling languages against requirements of MES. Unfortunately, this comparison is only presented as a table without additional argumentation. Furthermore, the authors introduce a new modeling notation which combines UML diagrams and a restricted version of BPMN to maintain ease of use. Execution of machine functions (e.g. in the form of skills) is not considered. 

The authors of \cite{ZSL_AProposalofBPMN_2011} provide an early proposition to extend BPMN to make it more suitable for manufacturing processes. The proposition includes a new task type \emph{Manufacturing Task}, new gateways to represent materials, additional flows as well as to model machines and parts. The approach is focused on graphically modeling processes and doesn't consider execution at all. 
In our approach, we use standard BPMN Service Tasks and extend BPMN with regard to the information of capabilities and skills which is stored in such a task. Apart from that, no additional extensions are required for our approach, even though the extension of \cite{ZSL_AProposalofBPMN_2011} may provide a better visual aid for users in modeling and reading diagrams.

The work of \cite{KVL_DatadrivenWorkflowManagementby_2019} is based on the IoT framework \emph{Eclipse Arrowhead}\footnote{https://www.arrowhead.eu/eclipse-arrowhead/} and the approach is very similar to ours. The authors recognize the distinction between an upper-level orchestration language and a lower-level behavior specification. 
While the authors use BPMN for the former just like we do, they use Colored Petri Nets for lower-level behavior specification. We model the behavior of individual skills using state machines. 
The approach of \cite{KVL_DatadrivenWorkflowManagementby_2019} does not feature a semantic description of functions. Furthermore, the lower-level behavior is implemented in classes that are directly referenced within a BPMN workflow, resulting in tight coupling between orchestration and implementation. In our approach, we use one delegation class that can be used to invoke skills in general. Additionally, we allow reuse of BPMN processes as they are bound to capabilities and not specific skills (see Section \ref{sec:approach}).

The contribution presented in \cite{LiFr_Towardsuserorientedprogrammingof_2019} introduces a process modeling approach that combines skill-based engineering with user-oriented workflow modeling. The authors base their approach on BPMN and introduce additional elements such as assets, special relations and actions. The approach allows to combine automated skill execution with manual operations that require visual instructions. However, with changes to the flow behavior, their adaptions of the BPMN model semantics are quite extensive.
Furthermore, the approach presented in \cite{LiFr_Towardsuserorientedprogrammingof_2019} has the same drawback as the approach presented in \cite{KVL_DatadrivenWorkflowManagementby_2019}: Skills are directly added into the workflow making it executable only for one particular plant setup. Reconfiguration of the plant would require to re-model workflows making the approach less suited for highly flexible production.

The authors of \cite{LiFr_Towardsuserorientedprogrammingof_2019} have further developed their approach \cite{SLF+_AgenericApproachfor_2020}. In \cite{SLF+_AgenericApproachfor_2020}, a more detailed orchestration process is presented. Nevertheless, this approach also does not distinguish between an abstract model of functions (capabilities) and concrete implementations (skills), which results in the aforementioned tight coupling between a workflow and skills. While there seems to be a possibility to have multiple instances for one type of a skill, this doesn't allow for the same level of abstraction that a distinction in capabilities and skills offer, e.g. because in \cite{SLF+_AgenericApproachfor_2020} all skills use the same technology (i.e. OPC UA). 
In our approach, one capability might be executable using multiple skills implemented in various technologies. We currently support web services and OPC UA, but the distinction of capabilities and skills allows to easily extend this in the future.
Besides the orchestration process, another interesting contribution of \cite{SLF+_AgenericApproachfor_2020} is a list of challenges in the context of skill-based orchestration. We considered these challenges as requirements for our own approach.

Summarizing related work, it can be stated that there is currently no other approach that clearly distinguishes between an abstract description of functions (i.e., capabilities) and executable implementations (i.e., skills) for modeling production processes. Furthermore, many of the existing approaches are based on custom modeling languages or highly customized variants of existing languages.
Our approach is based on a clear separation between capabilities (at modeling time) and skills (at execution time). In addition, we use BPMN in a standard-compliant way and thus ensure integration of skill orchestration with other (IT) processes such as messaging, error handling, and the use of other tasks (e.g., user tasks).

\section{Summary \& Outlook}
\label{sec:summary}
This contribution presented an approach to model general production recipes in a reusable manner and transform them to control recipes in order to execute them automatically with changing manufacturing plant configurations. 
First, a justified selection for a modeling language that can be used to model and execute manufacturing processes was made. BPMN was chosen as it provides powerful modeling abilities, a high degree of formalism and an extensive tool support.
Afterwards, we presented our approach to model manufacturing processes using capabilities and execute them using skills. 
Standard-compliant extensions to a BPMN modeling tool as well as an execution engine were made. This way, skill execution can be seamlessly combined with error handling, conventional tasks such as manual / user tasks, sending or receiving notifications or any kind of interaction with other web services.
Finally, an evaluation was presented in which we demonstrated two key points: 
First, modeling and storing production processes in the form of \emph{capability processes} and transforming them to \emph{skill processes} only for execution allows for a good abstraction of general production recipes and current plant layout.
Second, BPMN engines are a promising and robust means to execute skill processes that integrate with other automated and also manual tasks. 

\smallskip
One current limitation of our approach is the fact that in cases where there are multiple skills for one capability, a user has to manually choose a skill to be used in process execution. This decision could be done by a scheduling / optimization algorithm and this is something we will work on in our future research. There are existing contributions to capability-based scheduling \cite{KKZ_Capabilitybasedplanningandscheduling_2014}, but they typically focus on machine allocation or optimization of certain goals (e.g., completion time) and require a detailed production plan. 
We plan to integrate such scheduling approaches with more general planning solutions that can cope with more abstract descriptions and don't require a detailed plan. Initial steps for such an approach are presented in \cite{ESA+_RoadmaptoSkillBased_2019}.

Regardless of whether capability processes are modeled manually or created by a planning algorithm, they should be checked for correctness before execution. Such checks can be done using formal verification of BPMN processes and are another topic of future research. A recent contribution providing a comprehensive overview on existing BPMN verification approaches along with a powerful approach that covers both BPMN processes and collaboration is presented in \cite{CFP+_Aformalapproachfor_2021}. 
Such approaches could be extended for capability and skill processes.
First of all, approaches like the one presented in \cite{CFP+_Aformalapproachfor_2021} typically focus on checking the \emph{soundness} of a process model without checking the internal behavior of its tasks. In our approach, the behavior of every skill is modeled with a state machine so that in addition to the overall process model, skill behavior could also be considered in verification.
In addition to that, rules that are specific to a use case (e.g., that a certain skill must always be invoked after another one) need to be taken into account. How such rules can be specified by users without expertise in formal verification is another interesting topic for future research.

\bibliographystyle{IEEEtran}
\bibliography{references} 

\end{document}